# Title

COLD-CI: A large-scale very high-resolution label polygon dataset for cocoa and non-cocoa classification in Côte d'Ivoire

# Authors

Kasimir Orlowski[1,2,*], Michele Meroni[3], Astrid Verhegghen[4,5], Filip Sabo[6], Caroline Winchester[7], Felix Rembold[6,*], Martina Schneider[7]

[1]Faculty of Geo-Information Science and Earth Observation (ITC), University of Twente, Enschede, The Netherlands.
[2]FINCONS S.P.A., Vimercate, Italy.
[3]Seidor Consulting, Barcelona, Spain.
[4]ARHS Developments Italia S.R.L., Milan, Italy.
[5]Flemish Institute for Technological Research (VITO), Mol, Belgium.
[6]European Commission, Joint Research Centre (JRC), Ispra, Italy.
[7]World Resources Institute (WRI), Washington, DC, USA.

*Corresponding authors: kasimir.orlowski@utwente.nl; felix.rembold@ec.europa.eu

# Abstract

Spatially explicit information on cocoa cultivation is essential for land-use planning, deforestation monitoring, environmental assessment, and supply-chain analysis. Although several cocoa map products exist, their underlying reference data are often not publicly available, limiting transparency and methodological benchmarking. Here, we present a large-scale, very high-resolution **co**coa and non-cocoa **l**abel polygon **d**ataset for **C**ôte d'**I**voire (COLD-CI), covering the main cocoa-producing regions as well as contrasting non-cocoa landscapes.

COLD-CI consists of 123,736 vector polygons corresponding to a total labelled area of 5,996 km², including 58,107 cocoa polygons (1,788 km²) and 65,629 background polygons (4,208 km²). Polygon label candidates were first generated through conservative automated filtering of polygons from the West Africa Cocoa dataset and the combination of multiple external thematic datasets. These candidates were subsequently refined and complemented through systematic visual interpretation, manual correction, and digitisation using very high-resolution (0.5 m) satellite imagery. The resulting label polygons capture cocoa planted areas and associated fine-scale internal heterogeneity, as well as a wide range of non-cocoa land-cover types.

Independent validation using field-based and expert photointerpreted reference data from the Copernicus4GEOGLAM validation dataset indicated an overall agreement of 99%, with producer's and user's accuracy exceeding 98% for both cocoa and background classes. COLD-CI is released as a vector dataset with associated metadata to support transparent benchmarking, model development, and validation across a wide range of spatial resolutions.

# Background & Summary

Cocoa production is a major agricultural activity in West Africa and is closely associated with land-use change and forest conversion[1–3]. Information on the location and extent of cocoa cultivation is therefore important for applications such as land-use planning, deforestation monitoring, environmental assessment, and supply-chain analysis[1]. In particular, cocoa mapping is of prominent interest for the EU Regulation on deforestation-free supply chains (EUDR), which aims to ensure that key commodity supply chains are deforestation-free after the cut-off date of 31 December 2020[4].

Several Earth Observation–based cocoa mapping products have been developed at cross-country, national, and sub-national scales, particularly for major production areas in Côte d'Ivoire, Ghana, and Cameroon[1,5–10]. These products demonstrate the potential of Earth Observation (EO) data to map cocoa cultivation across large areas.

However, in many cases the reference data used for model development and evaluation are not publicly available, as they originate from private-sector inventories or proprietary field surveys. This limits transparency, model benchmarking, and general reuse. Publicly available cocoa reference datasets remain limited in number, spatial coverage, and label completeness. For Côte d'Ivoire and Cameroon, only small datasets are publicly available, including georeferenced cocoa plot inventories and drone imagery surveys covering tens to a few hundred cocoa parcels[11–14], while dedicated extensive EO cocoa label datasets remain absent. For Ghana, a larger dataset covering more than 20,000 cocoa polygons has been released recently[15]. In these datasets, background (non-cocoa) labels are generally absent or limited in extent and class diversity, constraining their direct use for classification model development and systematic benchmarking.

At the regional scale, the West African Cocoa (WAC) dataset[16] provides extensive polygon delineations of individual farmer cocoa plot locations across Côte d'Ivoire and Ghana. However, it was not designed for direct use in EO applications: it lacks background (non-cocoa) reference polygons, and the cocoa polygons represent farm plot boundaries that frequently contain mixed land cover, including non-cocoa vegetation, open areas, and infrastructure, requiring extensive cleaning and quality control before they can be used for EO model development. As a result, there is currently no large-scale, publicly accessible cocoa and non-cocoa label dataset for Côte d'Ivoire that is ready for direct use in EO applications.

To address these limitations, we developed COLD-CI, a large-scale, very high-resolution label dataset for Côte d'Ivoire designed for EO applications. In contrast to existing datasets, COLD-CI includes label polygons for both cocoa and non-cocoa classes and is released as a publicly accessible resource to enable transparent reuse.

Label polygons produced in this work were generated by integrating multiple reference sources, including the WAC polygon dataset[16], external thematic maps, and manual photointerpretation supported by very high-resolution (VHR) satellite imagery. VHR imagery was used for polygon digitisation, verification, and systematic cleaning to reduce label uncertainty. The objective was to create a cleaned, quality-controlled derivative of the WAC dataset complemented by thematically diverse and spatially distributed background polygons.

To make efficient use of limited resources for this time-consuming process, an initial layer of label polygon candidates was generated automatically using pragmatic and conservative decision rules.

These rules prioritised label correctness and polygon purity for both cocoa and background classes, even at the expense of discarding potentially valid areas, reflecting a deliberate trade-off between retained spatial coverage and reduced photointerpretation burden.

Beyond its spatial coverage, COLD-CI is characterised by its scale and composition. It comprises 123,736 labelled polygons covering 5,996 $km^2$, including 58,107 cocoa polygons (1,788 $km^2$) and 65,629 background polygons (4,208 $km^2$), distributed across diverse landscapes. Label polygon generation followed a conservative two-stage workflow combining automated filtering, systematic grid-based visual interpretation, manual correction, and independent accuracy assessment. As the labels are released as vector polygons, they can be used with EO imagery across a wide range of spatial resolutions, from commercial VHR imagery (0.5 m) to freely available Sentinel-1/2 (10 m) and Landsat data (30 m).

COLD-CI covers multiple cocoa-producing regions of Côte d'Ivoire (Figure 1), where cocoa cultivation is concentrated in the southern and central parts of the country and is dominated by smallholder production systems[1,16]. The labelled polygons are distributed across multiple VHR satellite scene footprints within the national cocoa belt and encompass heterogeneous agricultural landscapes, including those dominated by tree crops (e.g., cocoa, rubber, cashew, oil palm, and coconut palm), planted forests, forest remnants, and annual crops. COLD-CI further covers a range of production systems, from monoculture plantations to mixed cropping and agroforestry systems.

In addition to the main cocoa belt, COLD-CI also includes labelled polygons located in northern Côte d'Ivoire to capture representative non-cocoa landscapes at the national scale, including cashew-dominated systems, tree savanna, dry forests, and other non-cocoa land uses (Figure 1). The inclusion of these northern regions increases the geographic and land-cover diversity represented in COLD-CI.

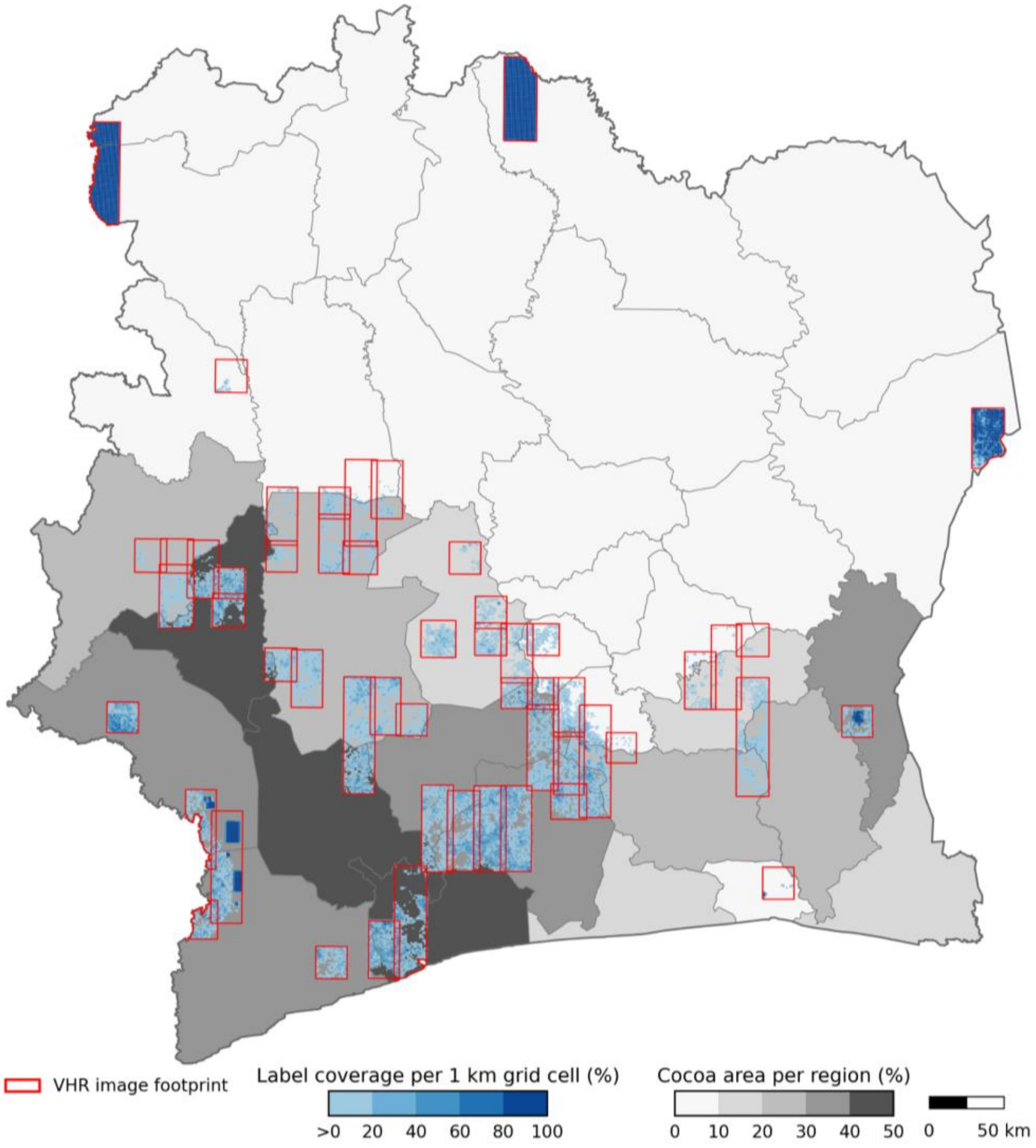

**Figure 1.** Spatial distribution and density of label polygons across Côte d'Ivoire. Label polygon area was aggregated to a regular 1 km grid and visualised as the percentage of grid-cell area covered by labels. Cocoa area per administrative unit (level-2) was extracted from the binary cocoa map from Kalischek et al. (2023)[1].

# Methods

## Input data

### Very high-resolution imagery

VHR optical satellite imagery from the Pléiades-1A and Pléiades-1B sensors was made available by the Copernicus Contributing Mission Mechanism in the context of the Copernicus Land Monitoring Service (CLMS) and was used as the primary visual reference for label generation. The imagery has a spatial resolution of 0.5 m and four spectral bands (red, green, blue, and near-infrared). In total, 138 Pléiades scenes were available across Côte d'Ivoire for the study period (2020-2021). Scenes affected by extensive cloud cover, severe atmospheric artefacts, or sensor noise were excluded, and overlapping acquisitions covering the same areas were filtered to avoid unnecessary redundancy. From the remaining scenes, 53 were selected for the construction of COLD-CI based primarily on overlap with available cocoa polygons (Figure 1).

The imagery was not acquired specifically for the construction of COLD-CI. Therefore, the spatial extent and number of scenes reflect the availability and suitability of Pléiades data purchased for a previous project[18]. As a result, COLD-CI does not provide continuous national coverage but represents the subset of areas for which high-quality VHR imagery and cocoa polygon data could be accessed.

### West Africa Cocoa polygon dataset

The primary cocoa data source was the WAC polygon dataset[16]. While the original dataset is proprietary, access was granted under a bilateral data-sharing agreement between the European Commission's Joint Research Centre and the World Resources Institute (WRI), in the framework of the European Commission's Sustainable Cocoa Initiative. The dataset contains polygon delineations of individual farmer cocoa plot locations compiled from multiple private-sector contributors participating in the Cocoa & Forest Initiative[17]. In total, the WAC dataset includes 839,242 cocoa polygons, of which 411,463 polygons are located in Côte d'Ivoire.

A subset corresponding to 102,519 of the WAC polygons was covered by the CLMS VHR imagery and retained for further processing. The WAC polygons represent individual farm plot boundaries rather than homogeneous canopy units and therefore often include mixed land cover such as non-cocoa vegetation, open areas, and infrastructure. As the polygons originate from heterogeneous data sources, they vary in positional accuracy, internal consistency, and annotation standards. Visual inspection against the VHR imagery further identified spatial misalignments and polygons that no longer corresponded to cocoa land cover, either due to land-use change or crop replacement since the time of data collection (e.g., conversion to oil palm or rubber), or due to misclassification in the source data.

## Copernicus4Geoglam validation dataset

The Copernicus4GEOGLAM validation dataset is a field- and photointerpretation-based reference dataset developed to support the independent validation of the national 2020 land-use and land-cover map of Côte d'Ivoire[18]. It was used in this study as an external source of reference data for assessing the quality of the COLD-CI label polygons.

The dataset comprises 7,515 sample points collected following a probabilistic two-stage stratified systematic random sampling design. Of these, 2,508 points were collected through field campaigns and 5,007 points through expert photointerpretation of VHR imagery.

At each sampling location, land cover was assessed within a 15 m radius for both field-based observations and VHR photointerpretation. Field data collection was carried out between November 2022 and March 2023.

Expert photointerpretation was conducted using VHR imagery from the Pléiades-1A and -1B, WorldView-1 and -3, and GeoEye sensors (0.5 m spatial resolution), with image acquisitions from 2020–2021.

## External thematic datasets

External thematic datasets were used as auxiliary sources to support automated filtering of cocoa polygons and to identify candidate background areas. These datasets provide thematic information on specific land-cover or crop classes but were not assumed to represent ground truth and were used conservatively to reduce obvious commission and omission errors prior to manual interpretation. Public datasets were selected based on relevance to Côte d'Ivoire, temporal proximity to the study period, accessibility, and transparent documentation of methods. All external thematic datasets were reprojected to a common coordinate reference system (WGS84; EPSG:4326) and internally aligned to a consistent 10 m spatial grid prior to integration.

Additionally, the following publicly available thematic datasets were initially considered but were not retained for candidate generation: the community models (cocoa, coffee, palm, rubber) released by Forest Data Partnership[19], the WorldCover and WorldCereal datasets from the European Space Agency[20,21], and the High Resolution 1 m Global Canopy Height dataset from Tolan et al. (2024)[22]. All datasets were screened through visual comparison with the VHR imagery but were excluded when their thematic accuracy, spatial consistency, or temporal alignment was insufficient to reliably support candidate generation, or when their inclusion was expected to increase manual correction effort without commensurate benefit.

### Cocoa indicator layers

Two national-scale cocoa indicator layers derived from high spatial resolution satellite imagery were used as auxiliary thematic sources. First, the cocoa class isolated from the BNETD Côte d'Ivoire Land-Cover 2020 map[7] with 23 thematic classes (hereafter referred to as the BNETD cocoa class) was used to provide a thematic indication of cocoa presence.

Second, a cocoa probability layer for Côte d'Ivoire produced by Kalischek et al. (2023)[1] was used to derive a high cocoa probability mask ($P \geq 0.65$), following the threshold identified by the authors as optimal based on a held-out validation dataset. This mask (hereafter referred to as the high cocoa probability mask) was used as an additional indicator of potential cocoa presence and complements the categorical information provided by the BNETD cocoa class.

### Background indicator layers

Closed-canopy oil palm plantations were obtained from the global smallholder and industrial oil palm map by Descals et al. (2021)[23], hereafter referred to as the oil palm layer. Closed-canopy coconut plantations were obtained from the global coconut plantation map by Descals et al. (2023)[24], hereafter referred to as the coconut layer. In addition, a mask of undisturbed forest was constructed (hereafter referred to as the forest layer). The forest layer was generated using a conservative multi-source approach. First, a forest extent mask was built combining the dense tree cover classes (classes 75-91 for terra firma and 195-211 for wetland forest) of the University of Maryland Global Land Cover and Land Use dataset (UMD GLC) 2019[25], the tropical tree extent probability dataset by Brandt et al. (2023)[26] thresholded at P ≥ 0.90, and the undisturbed forest class (class 10) of the JRC Tropical Moist Forest (TMF) dataset[27]. Areas where at least two of these datasets indicated tree extent or forest were kept to define the forest extent. In a second step, any area characterised by tree cover loss in the period 2001-2022 (as mapped by Hansen et al., 2013[28]) was masked out. In a third step, we applied a morphological operation (buffer in and out) to remove single pixels.

The cocoa probability layer of Kalischek et al. (2023)[1] was again used and thresholded to derive a low cocoa probability mask (P ≤ 0.02), hereafter referred to as the low cocoa probability mask. This mask was used as a conservative indicator of areas unlikely to contain cocoa and supported identification of candidate background areas not covered by the above background indicator layers, such as water bodies, bare soil, urban areas, grassland and savanna, and annual cropland or fallow fields.

VHR canopy height information at 0.5 m spatial resolution was obtained on the full extent of the VHR images through inference using the canopy height model (CHM) from Tolan et al. (2024)[22]. The CHM was generated by the extraction of features from a self-supervised model trained on Maxar VHR imagery, and the training of a dense prediction decoder against aerial Light Detection and Ranging (LIDAR) maps. To retrieve seamless mosaics of canopy height across images we predicted canopy height for tiles of 128 × 128 m (256 × 256 pixels). The selected tile size corresponds to the tile size that was used by Tolan et al. (2024)[22] during training of the CHM. The NIR band of our VHR images was disregarded during inference as the CHM was trained on RGB channels only. Information on canopy height was thresholded (≤ 0.025 m) to derive a fine-scale mask for bare and low vegetated areas, hereafter referred to as bare area mask. We selected this threshold based on empirical tests conducted in different areas across various VHR imagery to reliably identify bare soil and very low vegetation. The mask was applied within and immediately adjacent to candidate cocoa polygons to delineate fine-scale internal and boundary non-cocoa components, and was not used for the independent generation of background candidate areas.

## Label class definitions

Cocoa polygons represent cocoa planted area and its fine-scale internal heterogeneity rather than pure cocoa canopy or entire farm plot boundaries. Cocoa planted area is defined to include cocoa trees and co-occurring non-cocoa trees that form part of agroforestry systems, and may contain small non-woody gaps such as open spaces, herbaceous cover, or annual crops within mixed-cropping systems, which are typical for cocoa plots. The background class represents all non-cocoa land cover types (hereafter referred to as non-cocoa when discussing land cover in general).

# Label generation workflow

Label generation followed a two-stage workflow consisting of automated candidate generation and subsequent grid-based visual interpretation and manual correction (Figure 2). The objective of this workflow was to efficiently identify cocoa and background candidates and, through refinement and validation, convert them into cocoa and background label polygons while balancing spatial coverage, label quality, and manual workload.

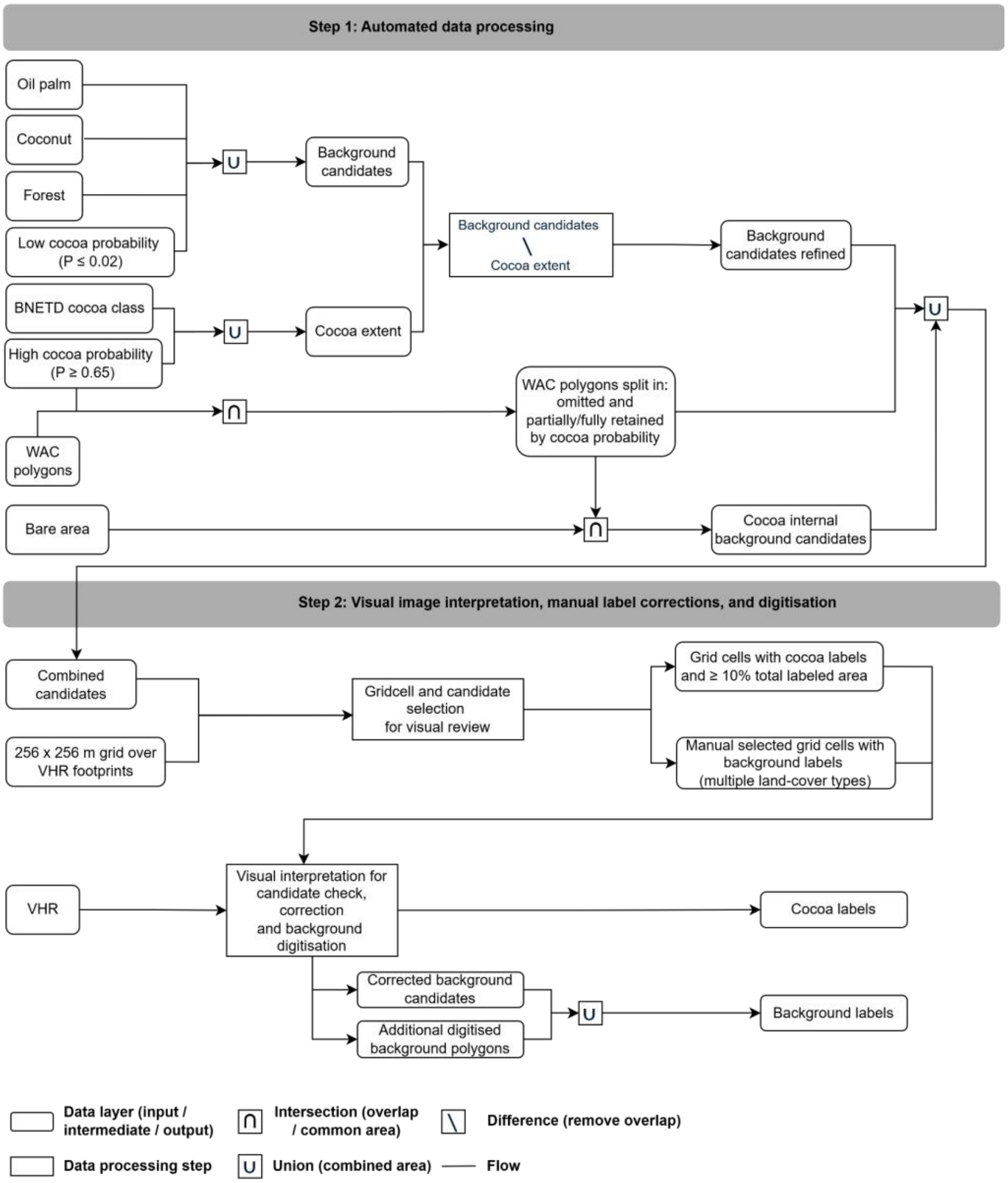

**Figure 2.** Overview of the label generation workflow. The flowchart illustrates the integration of multiple reference sources for automated candidate generation, followed by grid-based visual screening and manual correction to produce cocoa and background label polygons.

## Automated candidate generation

Automated candidate generation integrated three complementary information sources: (i) WAC cocoa polygons, (ii) cocoa indicator layers (BNETD cocoa class and high cocoa probability), and (iii) background indicator layers (oil palm, coconut, forest, low cocoa probability, and bare area).

As a first step, two derived layers were defined, a cocoa extent layer and a background candidate layer. The cocoa extent was defined by combining the BNETD cocoa class and the high cocoa probability mask. In parallel, all background indicators (except the bare area mask) were combined to define the background candidate layer.

The overlay of the cocoa extent with the background candidate layer revealed a small proportion of overlapping areas. Visual inspection of these overlap regions using the VHR imagery indicated that cocoa was present in some cases, reflecting residual classification uncertainty in the auxiliary layers. To favour background label purity, the background candidate layer was therefore refined by subtracting any area where the cocoa extent layer indicated cocoa presence. This conservative choice may also remove some true background areas but minimises the risk of cocoa contamination in background labels. The resulting intersection occasionally contained small background fragments. To reduce photointerpretation burden and avoid introducing small fragmented or noise-driven features, background candidate areas smaller than or equal to 100 m² were removed. Following this raster-based refinement, the background candidate raster was converted to vector polygons.

WAC polygons were used as the initial source of candidate cocoa areas. To support subsequent manual interpretation, a preliminary candidate delineation was automatically generated by intersecting the WAC polygons with the high cocoa probability mask. After this step, the overlapping portions of 92,418 WAC polygons were retained. In addition, 10,101 WAC polygons showing no spatial intersection with the high cocoa probability mask were kept and passed to manual interpretation rather than discarded automatically, to avoid omitting true cocoa areas that may fall outside the probability mask due to model omission errors.

The BNETD cocoa class was not used for automatic candidate delineation because reported omission rates are relatively high (30% based on independent validation[18]), and using it for this purpose would risk excluding substantial areas of true cocoa. In contrast, the cocoa probability model of Kalischek et al. (2023)[1] reports a recall of 90.9% for the cocoa class, making it more suitable for conservative candidate delineation where omission of true cocoa should be minimised.

In most cases, the intersection-based filtering effectively removed large non-cocoa areas within the WAC polygons (e.g. open areas, infrastructure, grassland) and improved boundary alignment of cocoa planted area.

We then intersected the CHM-derived bare area mask with the retained WAC polygons to identify fine-scale internal non-cocoa components (bare soil and very low vegetation) that were not resolved by the high cocoa probability mask due to its coarser spatial resolution. Only non-cocoa components with an area ≥ 200 m² were retained as background candidates for subsequent visual screening, to avoid labelling small non-woody gaps between cocoa tree stands as background, which are typical features of cocoa planted areas. The 200 m² threshold was selected based on visual inspection across cocoa plots and reflects the scale at which gaps can no longer be considered part of the cocoa planted area but instead constitute distinct non-cocoa land-cover features.

Lastly, the refined WAC and all background polygons were merged to form a combined candidate layer representing all areas to be screened during manual interpretation.

## Visual screening and manual correction

To facilitate systematic visual inspection of spatial polygons using VHR imagery, a regular spatial grid was overlaid across the study area. The grid served as a navigation framework to ensure consistent coverage during visual assessment of candidate geometries. A grid size of 256 × 256 m was selected to balance interpretability and spatial context. Cells were large enough to provide sufficient contextual information to support reliable discrimination of cocoa and non-cocoa land cover, while remaining small enough to allow efficient visual screening.

The gridding system resulted in a large number of grid cells to be examined (n=610,233). To optimise manual workload and focus the effort on information-rich cells, grid cells containing little labelled area were excluded from further analysis. Specifically, only grid cells containing at least 10% of labelled area were retained (n=174,162).

From this subset, all grid cells intersecting cocoa candidates were selected (n=107,902) to ensure that all potential cocoa areas were screened and that transitions between cocoa and non-cocoa land cover were represented. These transition zones are critical for downstream deep learning model development and evaluation because they define class boundaries and influence model behaviour at class transitions.

Furthermore, additional grid cells dominated by background candidates were purposively selected (n=66,260) to improve dataset completeness beyond cocoa-dominated landscapes. The majority were located in northern Côte d'Ivoire, targeting representative non-cocoa landscapes including tree savanna, dry forest, and cashew-dominated systems. The remainder were selected within the cocoa belt to capture contrasting land-cover types frequently co-occurring with cocoa, including forest patches, rubber, mango, and other tree crop plantations.

Within each selected grid cell, candidate cocoa and background geometries were visually inspected against the VHR imagery and manually corrected where necessary. Corrections included removal of remaining commission errors in both cocoa (n=24,687) and background (n=23,101) candidates, digitisation of additional background polygons (n=12,232), particularly along cocoa–non-cocoa boundaries, and removal of uncertain areas (n=3,340) from the label set when reliable class confirmation was not possible (e.g., due to cloud cover in the VHR imagery).

In COLD-CI, polygon boundaries were dissolved within each class to remove artificial internal boundaries introduced by intermediate geoprocessing steps (e.g. intersection, union, and clipping) or polygon boundary overlaps. This operation was applied to ensure that spatially contiguous regions belonging to the same class are represented as coherent geometries, independent of the original digitisation or preprocessing segmentation.

It is noted that, because a grid was used as a navigation framework during photointerpretation and only a subset of grid cells was selected for screening, COLD-CI was clipped to the screened grid cell, and candidate geometries in isolated grid cells may in some cases be partially truncated where they extend beyond grid boundaries.

# Data Records

COLD-CI is distributed through two complementary access pathways: an open-access Zenodo repository containing the validated background label polygons, VHR scene footprints, and associated documentation, and the full label polygon dataset, including cocoa polygons, available through WRI under a Data Sharing Agreement.

## Open repository files

The Zenodo repository (DOI: 10.5281/zenodo.20743810) contains the following openly accessible files: (i) background_labels.gpkg, containing validated non-cocoa label polygons and associated attributes; (ii) vhr_scene_footprints.gpkg, containing footprints of the VHR image scenes used during manual interpretation; and (iii) README.md, describing file contents, attribute definitions, coordinate reference system, and recommended usage. All vector files are provided in the WGS84 geographic coordinate reference system (EPSG:4326).

### Background label polygons

The file background_labels.gpkg contains 65,629 validated non-cocoa polygons covering 4,208 km². These polygons represent a broad range of non-cocoa land-cover types, such as forest, other tree crops, savanna, cropland, settlements, bare areas, and water bodies.

Each polygon record includes the following attributes:

- fid: unique polygon identifier
- class: categorical label (background)
- area_m2: polygon area in square metres
- sampling_region: sampling zone (cocoa_core or non_cocoa_region)

### VHR scene footprints

The file vhr_scene_footprints.gpkg contains polygons representing the footprints of the source VHR image scenes used during dataset construction.

Each polygon record includes the following attributes:

- fid: unique polygon identifier
- scene_id: unique scene identifier
- acquisition_year: acquisition year of the source imagery

## Documentation

The file README.md provides documentation for the openly accessible Zenodo repository files, including file contents, attribute definitions, coordinate reference systems, and recommended usage.

# Controlled-access full label dataset

The cocoa label polygons are derived from the West Africa Cocoa dataset and contain individual farmer plot locations. As these constitute sensitive data, they are not redistributed through the open Zenodo repository. The full COLD-CI label polygon dataset is instead available through WRI under a Data Sharing Agreement, as described in the Data Availability and Usage Notes sections.

This controlled-access release contains 123,736 validated label polygons covering 5,996 $km^2$ in total, comprising 58,107 cocoa planted-area polygons (1,788 $km^2$) and 65,629 background polygons (4,208 $km^2$).

The controlled-access dataset is provided in GeoPackage format and includes the following attributes:

- fid: unique polygon identifier
- class: categorical label (cocoa or background)
- area_m2: polygon area in square metres
- sampling_region: geographic sampling zone indicating polygons originating from the main cocoa-producing belt (cocoa_core or non_cocoa_region)

Representative examples of VHR imagery and corresponding label polygons across different landscape contexts are shown in Figure 3.

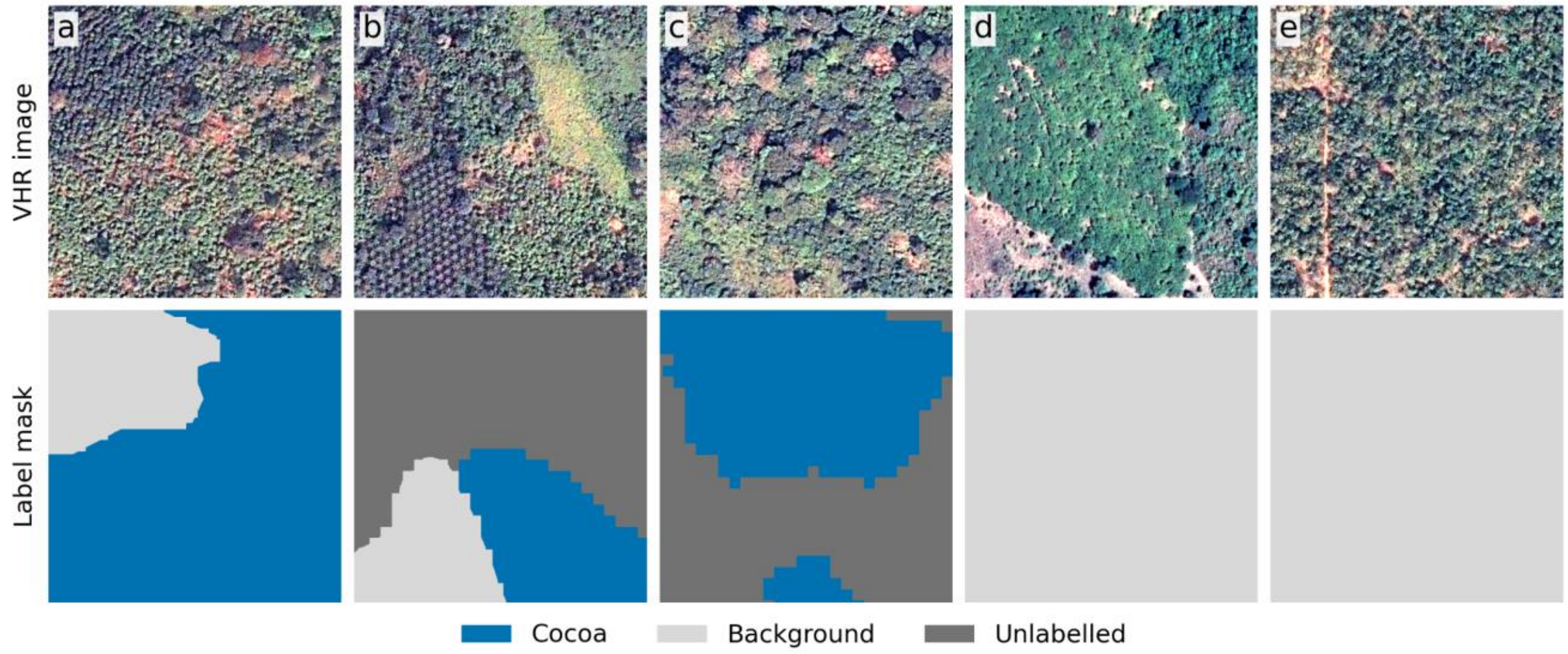

**Figure 3.** Representative examples of VHR imagery from Airbus Pléiades and corresponding label polygons across different landscape contexts in Côte d'Ivoire. Panels show (a) cocoa adjacent to a rubber plantation; (b) cocoa adjacent to a palm plantation; (c) shaded cocoa; (d) a non-cocoa cashew-dominated landscape in northern Côte d'Ivoire; and (e) a non-cocoa plantation forest landscape. Some boundary patterns reflect the influence of raster-based auxiliary layers and probability filtering used during label generation. Image data: © CNES (2020), distribution Airbus Defence and Space

# Technical Validation

Label quality was assessed through independent validation using reference points from the Copernicus4GEOGLAM validation dataset[18].

For this analysis, both field-based and expert photointerpreted reference samples were aggregated into binary classes (cocoa vs background). Field observations were restricted to samples classified as uniform in the original validation dataset, indicating that the observed land cover was considered homogeneous around the reference location. Each reference point was spatially intersected with the corresponding label polygon to extract the assigned class, which was then compared to the reference label. Reference points falling outside the spatial extent of the label dataset were excluded from the agreement assessment. After filtering, 464 reference points were retained for validation, comprising 178 cocoa and 286 background points.

Agreement between COLD-CI and the Copernicus4GEOGLAM reference points is summarised in Table 1 and 2. Overall accuracy was 99%. Producer's accuracy for both the cocoa and background was 99% as well, and user's accuracy was 98% for cocoa and 99% for background. Cohen's kappa was 0.98 and the F1 score for the cocoa class was 99% and 98% for background. These results provide an independent consistency check of the presented dataset.

**Table 1.** Confusion matrix comparing dataset labels with Copernicus4GEOGLAM reference points

| | Label cocoa | Label background | **Total** |
|---|---|---|---|
| Reference cocoa | 176 | 2 | **178** |
| Reference background | 3 | 283 | **286** |
| **Total** | **179** | **285** | **464** |

**Table 2.** Quantitative accuracy metrics for cocoa and background labels

| **Class** | **PA (%)** | **UA (%)** | **F1 (%)** | **Kappa** | **Accuracy (%)** |
|---|---|---|---|---|---|
| Cocoa | 99 | 98 | 99 | 98 | 99 |
| Background | 99 | 99 | 99 | | |

PA= Producer's accuracy, UA= User's accuracy

Using the same independent reference points, we additionally assessed whether the refinement workflow improved label purity relative to the original WAC polygons. User's accuracy for the cocoa class — defined as the proportion of reference points inside cocoa polygons carrying a cocoa reference label — increased from 90% for the original WAC polygons to 98% for the refined labels, indicating a reduction in cocoa label noise. Of the 225 reference points intersecting the original WAC cocoa polygons, 22 carried a background reference label, predominantly rubber (n=13), arable/fallow land (n=4), cashew (n=3), and degraded forest (n=2). The refinement workflow excluded the areas containing 19 of these 22 background-labelled points (86%) from the cocoa class, retaining only 3 residual commission errors: degraded forest (n=2) and cashew (n=1).

# Usage Notes

An anonymised version of the full COLD-CI label polygon dataset, including cocoa polygons derived from the WAC dataset, can be accessed through WRI under a Data Sharing Agreement. Staff of organisations with in-house expertise for data handling and analysis, a record of maintaining data confidentiality, and whose mandate is sufficiently distinct from farmer resettlement-related processes may qualify.

The Data Sharing Agreement acknowledges that the data will be considered as confidential information of WRI, securely stored and handled, and deleted once derivative works are completed. Permissible use cases include the development of derivative works which do not identify nor produce findings for individual cocoa plot locations. These derivative works and their supporting methodology documents must go through a peer review process, properly cite WRI's WAC technical note[16], and be made publicly available without charge for viewing and downloading under a Creative Commons CC BY 4.0 licence.

Users should note that COLD-CI includes polygons both from the main cocoa-producing regions and from regions outside the cocoa belt that were intentionally included to capture contrasting non-cocoa landscapes. The sampling_region attribute allows users to easily identify and filter polygons originating from the core cocoa regions ("cocoa_core") and from non-cocoa regions ("non_cocoa_region"). Depending on the target application and class balance requirements, users may choose to selectively sample or subsample polygons from each group when constructing training, validation, or evaluation datasets.

Users working with coarser-resolution imagery (e.g., Sentinel-2) should note that the label polygons resolve fine-scale internal non-cocoa components within cocoa plots based on VHR interpretation and extractions from the CHM. Depending on the target spatial resolution, minimum mapping unit and aggregation strategy, users may wish to merge small internal background patches into surrounding cocoa areas when resampling the labels to coarser grids to avoid artificial fragmentation or label noise.

# Data Availability

The open-access components of the COLD-CI dataset, including the non-cocoa background label polygons and VHR scene footprints, are available through Zenodo:

DOI: 10.5281/zenodo.20743810

Source imagery is not distributed due to licensing restrictions. COLD-CI label polygon dataset, including cocoa polygons derived from the WAC dataset, is
available through WRI for staff of organisations that meet certain criteria and sign a Data Sharing Agreement. Please reach out via WRI's Global Nature Watch request form to initiate the Data Sharing Agreement process:
https://globalnaturewatch.zendesk.com/hc/en-us/requests/new

# Funding

The authors received no funding for this work.

# References

1. Kalischek, N., Lang, N., Renier, C. et al. Cocoa plantations are associated with deforestation in Côte d'Ivoire and Ghana. Nat. Food 4, 384–393 (2023). https://doi.org/10.1038/s43016-023-00751-8

2. Renier, C. *et al.* Direct and indirect deforestation for cocoa in the tropical moist forests of Ghana. *Environ. Res.: Food Syst.* **2**, 025006 (2025). https://doi.org/10.1088/2976-601X/add01b

3. Singh, C. & Persson, U. M. Global patterns of commodity-driven deforestation and associated carbon emissions. Nat. Food 7, 138–151 (2026). https://doi.org/10.1038/s43016-026-01305-4

4. European Union. Regulation (EU) 2023/1115 of the European Parliament and of the Council of 31 May 2023 on the making available on the Union market and the export from the Union of certain commodities and products associated with deforestation and forest degradation and repealing Regulation (EU) No 995/2010. *Off. J. Eur. Union* L 150, 206 (2023). https://eur-lex.europa.eu/legal-content/EN/TXT/PDF/?uri=CELEX:02023R1115-20251226

5. Masolele, R. N. *et al.* Mapping the diversity of land uses following deforestation across Africa. *Sci. Rep.* **14**, 1681 (2024). https://doi.org/10.1038/s41598-024-52138-9

6. Moraiti, N., Mullissa, A., Rahn, E., Sassen, M. & Reiche, J. Critical assessment of cocoa classification with limited reference data: a study in Côte d'Ivoire and Ghana using Sentinel-2 and random forest model. Remote Sens. 16, 598 (2024). https://doi.org/10.3390/rs16030598

7. BNETD-CIGN. Note méthodologique de la carte d'occupation du sol de Côte d'Ivoire 2020 (2020). https://africageoportal.maps.arcgis.com/sharing/rest/content/items/76dc18767b89472eb89e8aa54e08a6c9/data

8. Numbisi, F. N., Van Coillie, F. M. B. & De Wulf, R. Delineation of cocoa agroforests using multiseason Sentinel-1 SAR images: a low grey level range reduces uncertainties in GLCM texture-based mapping. *ISPRS Int. J. Geo-Inf.* **8**, 179 (2019). https://doi.org/10.3390/ijgi8040179

9. Ashiagbor, G. et al. Pixel-based and object-oriented approaches in segregating cocoa from forest in the Juabeso-Bia landscape of Ghana. Remote Sens. Appl. Soc. Environ. 19, 100349 (2020). https://doi.org/10.1016/j.rsase.2020.100349

10. Abu, I. O., Szantoi, Z., Brink, A., Robuchon, M. & Thiel, M. Detecting cocoa plantations in Côte d'Ivoire and Ghana and their implications on protected areas. Ecol. Indic. 129, 107863 (2021). https://doi.org/10.1016/j.ecolind.2021.107863

11. Blaser-Hart, W. J. & Hart, S. The unrealised potential of agroforestry (climate mitigation potential analysis). University of Queensland institutional archive (2025). https://doi.org/10.48610/dda018c

12. Kouamé, I. K. et al. Supporting dataset for the article 'Maximizing tree diversity in cocoa agroforestry: taking advantage of planted, spontaneous, and remnant trees'. Zenodo

(2025). https://doi.org/10.5281/zenodo.15124220

13. Lescuyer, G. Inventaire des arbres dans 223 cacaoyères agroforestières au Cameroun. Harvard Dataverse (2024). https://doi.org/10.18167/DVN1/MGDIJU

14. Lammoglia, S. et al. High-resolution multispectral and RGB dataset from UAV surveys of ten cocoa agroforestry typologies in Côte d'Ivoire. Data Brief 55, 110664 (2024). https://doi.org/10.1016/j.dib.2024.110664

15. Centre for Remote Sensing and Geographic Information Services (CERSGIS). Reference dataset for land use change mapping in Ghana's cocoa landscape (2024–2025) (v2.0) [Data set]. Zenodo (2025). https://doi.org/10.5281/zenodo.16579443

16. Schneider, M., Winchester, C., Goldman, E. & Shao, Y. Mapping cocoa and assessing deforestation risk for the cocoa sector in Côte d'Ivoire and Ghana. World Resources Institute (2023). https://doi.org/10.46830/writn.21.00011

17. World Cocoa Foundation. Cocoa & Forests Initiative. Online resource (2026). https://worldcocoafoundation.org/programmes-and-initiatives/cocoa-and-forests-initiative

18. CLS & Terrasphere. *COPERNICUS4GEOGLAM Validation Note LandCover v2 Ivory Coast* (2024). https://africageoportal.maps.arcgis.com/sharing/rest/content/items/0072843eefe545b1b43b0c5b2ebec41a/data

19. Forest Data Partnership. Community models 2025a. GitHub repository (2025). https://github.com/google/forest-data-partnership/blob/main/models/model_2025a/README.md

20. Zanaga, D. et al. ESA WorldCover 10 m 2020 v100. Zenodo (2021). https://doi.org/10.5281/zenodo.5571936

21. Van Tricht, K. et al. WorldCereal: a dynamic open-source system for global-scale, seasonal, and reproducible crop and irrigation mapping. Earth Syst. Sci. Data 15, 5491–5515 (2023). https://doi.org/10.5194/essd-15-5491-2023

22. Tolan, J. et al. Very high resolution canopy height maps from RGB imagery using self-supervised vision transformer and convolutional decoder trained on aerial lidar. Remote Sens. Environ. 300, 113888 (2024). https://doi.org/10.1016/j.rse.2023.113888

23. Descals, A. et al. High-resolution global map of smallholder and industrial closed-canopy oil palm plantations. Earth Syst. Sci. Data 13, 1211–1231 (2021). https://doi.org/10.5194/essd-13-1211-2021

24. Descals, A. et al. High-resolution global map of closed-canopy coconut palm. Earth Syst. Sci. Data 15, 3991–4010 (2023). https://doi.org/10.5194/essd-15-3991-2023

25. Hansen, M. C. et al. Global land use extent and dispersion within natural land cover using Landsat data. Environ. Res. Lett. 17, 034050 (2022). https://doi.org/10.1088/1748-9326/ac46ec

26. Brandt, J., Ertel, J., Spore, J. & Stolle, F. Wall-to-wall mapping of tree extent in the tropics with Sentinel-1 and Sentinel-2. Remote Sens. Environ. 292, 113574 (2023). https://doi.org/10.1016/j.rse.2023.113574

27. Vancutsem, C. et al. Long-term (1990–2019) monitoring of forest cover changes in the

humid tropics. Sci. Adv. 7, eabe1603 (2021). https://doi.org/10.1126/sciadv.abe1603

28. Hansen, M. C. et al. High-resolution global maps of 21st-century forest cover change. Science 342, 850–853 (2013). https://doi.org/10.1126/science.1244693